\documentclass{ws-ijmpe}
\usepackage[super,compress]{cite}

\begin{document}

\markboth{M. Thoennessen}{2014 Update of the Discoveries of Isotopes}

\catchline{}{}{}{}{}

\title{2014 UPDATE OF THE DISCOVERIES OF NUCLIDES}

\author{\footnotesize M. THOENNESSEN}

\address{National Superconducting Cyclotron Laboratory and \\
Department of Physics \& Astronomy \\
Michigan State University\\
East Lansing, Michigan 48824, USA\\
thoennessen@nscl.msu.edu}

\maketitle

\begin{history}
\received{Day Month Year}
\revised{Day Month Year}
\end{history}

\begin{abstract}
The 2014 update of the discovery of nuclide project is presented. Only six new nuclides were observed for the first time in 2014 while the assignments of seventeen other nuclides were revised. In addition, for another fourteen nuclides the laboratories where they were discovered were reassigned.
\end{abstract}


\ccode{PACS numbers: 21.10.-k, 29.87.+g}


\section{Introduction}

This is the second update of the isotope discovery project which was originally published in a series of papers in Atomic Data and Nuclear Data Tables beginning in 2009 with the discovery of the cerium isotopes \cite{2009Gin01} and was completed with the publication of the discovery of dysprosium, holmium, erbium, thulium, and ytterbium isotopes in 2013 \cite{2013Fry01}. Subsequently, comprehensive overviews were  published in 2012 and 2013 in Nuclear Physics News \cite{2012Tho03} and Reports on Progress in Physics \cite{2013Tho02}, respectively. The present journal published the first update at the beginning of last year \cite{2014Tho01}. 

\section{New discoveries in 2014}

In 2014, the discoveries of only six new nuclides were reported in refereed journals. One proton-unbound resonance, one proton-rich nucleus, two neutron-rich nuclei close to the neutron dripline, one proton-rich $\alpha$-emitting nucleus, and one transuranium nuclide at the end of an $\alpha$ decay chain of a superheavy nuclide were discovered and are listed in Table \ref{2013Isotopes}.

\begin{table}[pt]
\tbl{New nuclides reported in 2014. The nuclides are listed with the first author, submission date, and reference of the publication, the laboratory where the experiment was performed, and the production method (SB = secondary beams, PF = projectile fragmentation, FE = fusion evaporation). \label{2013Isotopes}}
{\begin{tabular}{@{}llcclc@{}} \toprule 
Nuclide(s) & Author & Subm. Date & Ref. & Laboratory & Type \\ \colrule
$^{15}$Ne&  F. Wamers et al.  & 1/17/2014 &  \refcite{2014Wam01} &  GSI   & SB  \\
\parbox[t]{2.5cm}{\raggedright $^{77}$Co, $^{80}$Ni \vspace*{0.1cm}}&  Z.Y. Xu et al.  & 3/28/2014 &  \refcite{2014Xu01} &  RIKEN  & PF  \\
$^{164}$Ir&  M.C. Drummond et al.  & 4/15/2014 &  \refcite{2014Dru01} &  Jyv\"askyl\"a  & FE \\
$^{205}$Ac&  Z.Y. Zhang et al.  & 12/1/2013 &  \refcite{2014Zha01} &   Lanzhou   & FE \\
$^{266}$Lr&   J. Khuyagbaatar et al. & 2/22/2014 &  \refcite{2014Khu01}  &  GSI   & FE \\
\botrule
\end{tabular}}
\end{table}

The discovery of $^{15}$Ne was reported by F. Wamers {\it et al.} in ``First Observation of the Unbound Nucleus $^{15}$Ne'' \cite{2014Wam01}. A 500 MeV $^{17}$Ne secondary beam - produced from a primary $^{20}$Ne delivered by the heavy-ion synchrotron SIS at GSI - bombarded a beryllium target to form $^{15}$Ne in a two-neutron knockout reaction. The $^{13}$O fragment was measured in coincidence with two protons after passing through the magnetic field of a large-gap dipole magnet. ``The $^{15}$Ne ground state was found to be unstable towards $^{13}$O + p + p decay by 2.522(66) MeV, which corresponds to an atomic mass excess of ME($^{15}$Ne) = 40.215(69) MeV.''

In the paper ``$\beta$-Decay Half-Lives of $^{76,77}$Co, $^{79,80}$Ni, and $^{81}$Cu: Experimental Indication of a Doubly Magic $^{78}$Ni'' the nuclides $^{77}$Co and $^{80}$Ni were reported for the first time \cite{2014Xu01}. A 345 MeV/nucleon $^{238}$U from the RIKEN cyclotron accelerator complex was incident on a beryllium target and projectile fragments were analyzed with the BigRIPS separator and the ZeroDegree Spectrometer. The fragments were then implanted in the highly segmented beam stopper, wide-range active silicon
strip stopper array for beta and ion detection (WAS3ABi). ``The half-lives of 20 neutron-rich nuclei with Z = 27$-$30 have been measured at the RIBF, including five new half-lives of $^{76}$Co(21.7$^{+6.5}_{-4.9}$ ms), $^{77}$Co(13.0$^{+7.2}_{-4.3}$ ms), $^{79}$Ni(43.0$^{+8.6}_{-7.5}$ ms), $^{80}$Ni(23.9$^{+26.0}_{-17.2}$ ms), and $^{81}$Cu(73.2$\pm$6.8 ms).''

The first observation of $^{164}$Ir was published by Drummond {\it et al.} in the paper entitled ``$\alpha$ decay of the $\pi h_{11/2}$ isomer in $^{164}$Ir.'' \cite{2014Dru01} An isotopically enriched $^{92}$Mo target was bombarded with $^{78}$Kr beams of 428, 435 and 450 MeV from the K120 cyclotron at Jyv\"askyl\"a. The recoils from the fusion evaporation reaction $^{92}$Mo($^{78}$Kr,p5n)$^{164}$Ir were separated with the gas-filled separator RITU and identified in the GREAT spectrometer. ``The yield of approximately 100 $^{164}$Ir proton-decay events obtained in the present work allowed the half-life to be determined with improved precision. A half-life of 70$\pm$10 μs was obtained by using the method of maximum likelihood.'' Drummond {\it et al.} did not consider the observation a new discovery quoting previous results published in conference proceedings.\cite{2001Ket02,2002Mah01}

Zhang {\it et al.} published the first observation of $^{205}$Ac in ``$\alpha$ decay of the new neutron-deficient isotope $^{205}$Ac''  \cite{2014Zha01}. A $^{169}$Tm target was bombarded with a 198 MeV $^{40}$Ca beam from the Sector-Focusing Cyclotron of the Heavy Ion Research Facility in Lanzhou and $^{205}$Ac in the 4n fusion evaporation reaction. $^{205}$Ac were separated with the gas-filled recoil separator SHANS (Spectrometer for Heavy Atoms and Nuclear Structure) and implanted in a position sensitive silicon detector which also detected the subsequently emitted $\alpha$ particles. ``The $\alpha$ decay energy and half-life of $^{205}$Ac were determined to be 7.935(30) MeV and 20$^{+97}_{-9}$ ms, respectively.''

 Khuyagbaatar {\it et al.} described the discovery of $^{266}$Lr in the paper ``$^{48}$Ca + $^{249}$Bk Fusion Reaction Leading to Element Z = 117: Long-Lived $\alpha$-Decaying $^{270}$Db and Discovery of $^{266}$Lr'' \cite{2014Khu01}. $^{40}$Ca beams at 254 and 258 MeV bombarded $^{249}$Bk$_2$O$_3$ targets at the gas-filled Trans Actinide Separator and Chemistry Apparatus (TASCA) at GSI to form $^{294}$117 in the fusion evaporation reaction $^{249}$Bk($^{48}$Ca,3n). $^{266}$Lr was identified following a sequence of 7 $\alpha$ decays by its spontaneous fission. ``A hitherto unknown $\alpha$-decay branch in $^{270}$Db (Z = 105) was observed, which populated the new isotope $^{266}$Lr (Z = 103).''

\section{Changes of prior assignments}

In addition to the six new nuclides discovered in 2014 some of the previous assignments were reevaluated based on feedback from other researchers and new results. 

The discovery of $^{218}$At had been attributed\cite{2013Fry03} to the work by Karlik and Bernert \cite{1943Kar01} published in 1943. However, in an overview paper in 2010 Thornton and Burdette \cite{2010Tho01} pointed out that Hulubei and Cauchois reported evidence for the observation already in 1939. In their paper ``Spectres de l'emission propre ondulatoire du radon et de ses d\'eriv\'es. Raies attribuables \`a l'\'el\'ement 85'' Hulubei and Cauchois \cite{1939Hul02} had reported the observation of three X-rays which were close to the predicted values for eka-iodine (astatine). These X-rays were observed from a $^{222}$Rn sample and the astatine  lines could only originate from $^{218}$At populated either by $\alpha$ and $\beta$-decay through $^{218}$Po or by $\beta$ and $\alpha$-decay through $^{222}$Fr. ``Ces coincidences font penser que l'\'el\'ement 85 est peut-\^etre pr\'esent parmi les produits de d\'esint\'egration du radon'' (These coincidences suggest that the element 85 may be present among the radon decay products).\cite{1939Hul02}

The acceptance of the Hulubei and Cauchois discovery of $^{218}$At places the discovery of this nuclide before the discovery of the element astatine. The discovery of astatine has been credited \cite{1947Pen01} to Corson, MacKenzie and Segr\`e for their observation of the $\alpha$-decay of $^{211}$At in 1940.\cite{1940Cor03,1940Cor01} One can argue that Hulubei and Cauchois should not deserve credit for the discovery of the element because they lacked chemical proof.\cite{2010Tho01} 

The criteria for the discovery of a nuclide differ from the criteria for the discovery of an element so it is not inconsistent or contradictory that a nuclide of astatine was discovered before the element astatine. Similarly, nuclides of plutonium and californium have been discovered prior to the discovery of the respective elements.\cite{2013Fry05}

The second reassignment involves $^{12}$Li which was initially assigned to Yu. Aksyutina {\it et al.}\cite{2008Aks01} In this work, $^{12}$Li was populated in the reaction $^1$H($^{14}$Be,2pn)$^{12}$Li simultaneously with the discovery of $^{13}$Li which was produced in the one-proton knockout reaction. However, it was recently pointed out\cite{2013Koh01} that the $^{12}$Li data could have been contaminated by misidentified low energy two-neutron decay events from $^{13}$Li invalidating the extracted scattering length for $^{12}$Li. This interpretation is also supported by the analysis of two-neutron events from the decay of $^{26}$O which was performed with the same setup as the $^{12}$Li experiment.\cite{2013Cae01} Thus the discovery of $^{12}$Li is now credited to the 2010 paper ``First observation of excited states in $^{12}$Li by Hall {\it et al.}\cite{2010Hal01} A secondary 53.4 MeV/u $^{14}$B beam from the Coupled Cyclotron Facility at Michigan State University was used to bombard a beryllium target producing $^{12}$Li in a two-proton removal reaction. The Modular Neutron Array was used to measure neutrons in coincidence with the $^{11}$Li fragments. Resonances in $^{12}$Li were reconstructed from the invariant mass: ``Two excited states at resonance energies of 250$\pm$20 keV and 555$\pm$20 keV were observed for the first time.''

The assignments of fifteen nuclides were revised because the originally assigned papers corresponded to conference proceedings. 

The credit for the discovery of $^{79,80}$Zn by G. Rudstam {\it et al.} \cite{1981Rud01} and $^{71,93,94}$Br  by B. Vosicki {\it et al.} \cite{1981Vos01} were given\cite{2012Gro01} to papers published in Nuclear Instruments and Methods as part of the proceedings of the 10$^{th}$ International Conference on Electromagnetic Isotope Separators and Techniques Related to their Applications (EMIS) in 1980.\cite{1981EMI01} 

$^{79}$Zn was first reported in a referee publication in 1986 by B. Ekstr\"om {\it et al.} in ``Decay Properties of $^{75-80}$Zn and Q$_\beta$-Values of Neutron-Rich Zn and Ga Isotopes.'' \cite{1986Eks01} Fission fragments were separated with the OSIRIS ISOL facility at Studsvik. Gamma-rays were measured with a coaxial Ge detector and several spectra were recorded consecutively. ``The current multispectrum studies yielded half lives of (1.0$\pm$0.1)s and (0.53$\pm$0.05)s for $^{79}$Zn and $^{80}$Zn, respectively.'' Although they also reported a half-life measurement for $^{80}$Zn they do not get credit for the discovery because two months earlier R.L. Gill {\it et al.} had submitted their results in ``Half-Life of $^{80}$Zn: The First Measurement for an r-Process Waiting-Point Nucleus.'' \cite{1986Gil01} in 1986. Neutrons from the high-flux beam reactor HFBR at Brookhaven National Laboratory irradiated a $^{235}$U target and the fission fragments were separated with the mass separator TRISTAN. Time-sequential $\gamma$-ray spectra as well as $\gamma$-$\beta$ coincidence spectra were recorded. ``Analysis of the time-sequential spectra identified several $\gamma$ rays as belonging to the $^{80}$Zn decay. Strong $\gamma$ rays of 713 and 715 keV were used to
determine a $^{80}$Zn half-life of 0.55$\pm$0.02 s.''

E. Hagberg {\it et al.} discovered $^{71}$Br in 1982 as reported in ``The decay of a new nuclide: $^{71}$Br.'' \cite{1982Hag01} A 132 MeV $^{35}$Cl beam bombarded a natural calcium target and the evaporation residues were separated with the Chalk River on-line isotope separator.  Sequential $\gamma$-ray spectra were recorded with the help of a small cassette tape-transport system. ``The decay of mass-separated samples of the previously unknown nuclide $^{71}$Br have been investigated by means of the Chalk River on-line isotope separator. Eleven $\gamma$-rays were assigned to the decay of this nuclide, and its half-life was measured to be 21.4$\pm$0.6 s.''

$^{93}$Br and $^{94}$Br were first observed by K.-L. Kratz {\it et al.} in the 1988 paper ``Onset of Deformation in Neutron-Rich Krypton Isotopes.'' \cite{1988Kra01} Fission fragments were analyzed with the mass separators OSTIS and ISOLDE at Grenoble and CERN, respectively. Delayed neutrons, $\beta$-particles and $\gamma$-rays were measured in order to identify the nuclides. The half-lives for $^{93}$Br and $^{94}$Br were listed in a table as 102$\pm$10 ms and 70$\pm$20 ms, respectively.

The discovery of $^{166}$Eu by A. Osa {\it et al.} \cite{2008Osa01} was published in Nuclear Instruments and Methods B as part of the proceedings of the 15$^{th}$ EMIS conference in 2007.  It is now credited to ``Discovery and cross-section measurement of neutron-rich isotopes in the element range from neodymium to platinum with the FRS'' by J. Kurcewicz {\it et al.} \cite{2012Kur01} A 1 GeV/u $^{238}$U beam from the SIS-18 synchrotron at GSI bombarded a beryllium target in front of the projectile Fragment Separator FRS. The fragments were identified event-by-event by their time-of-flight, energy-loss and magnetic rigidity. `` The discovery of $^{166}$Eu was not specifically mentioned because it was assumed to be known.\cite{2013May01}

The discovery of $^{216-219}$Pb and $^{219-223}$Bi had been credited\cite{2013Fry02} to a 2009 paper by Alvarez-Pol {\it et al.} \cite{2009Alv01} which was part of the topical issue of the 5$^{th}$ International Conference on Exotic Nuclei and Atomic Masses (ENAM) in 2008. It has now been reassigned to a paper by the same group which was published a year later. \cite{2010ALv01}

Finally, for fourteen nuclides the laboratories where they had been discovered were revised. The assignments of early papers from the Institute of Physical and Chemical Research in Tokyo were incorrectly assigned to the University of Tokyo instead of RIKEN.\cite{2012Nys01,2012Par01,2010Fri01,2012Rob01} A total of 12 isotopes were reassigned ($^{92}$Y \cite{1940Sag01}, $^{89}$Zr \cite{1938Sag02}, $^{92,94}$Nb \cite{1938Sag03}, $^{91,99}$Mo \cite{1938Sag02}, $^{102}$Rh \cite{1941Min01}, $^{185,187}$W \cite{1940Min01}, $^{186,188}$Re \cite{1939Sin01}, and $^{237}$U \cite{1940Nis01}). Also,  the assignment\cite{2013Fry01} of $^{169}$Ho was changed from Tokyo to the Japan Atomic Energy Research Institute at Tokai \cite{1963Miy01}. In addition, the assignment\cite{2012May01} of $^{141}$La was changed from Argonne National Laboratory to Chicago as the experiment was performed at the Chicago cyclotron.\cite{1951Kat01}

\section{Status at the end of 2014}

\begin{figure}[pt] 
\centerline{\psfig{file=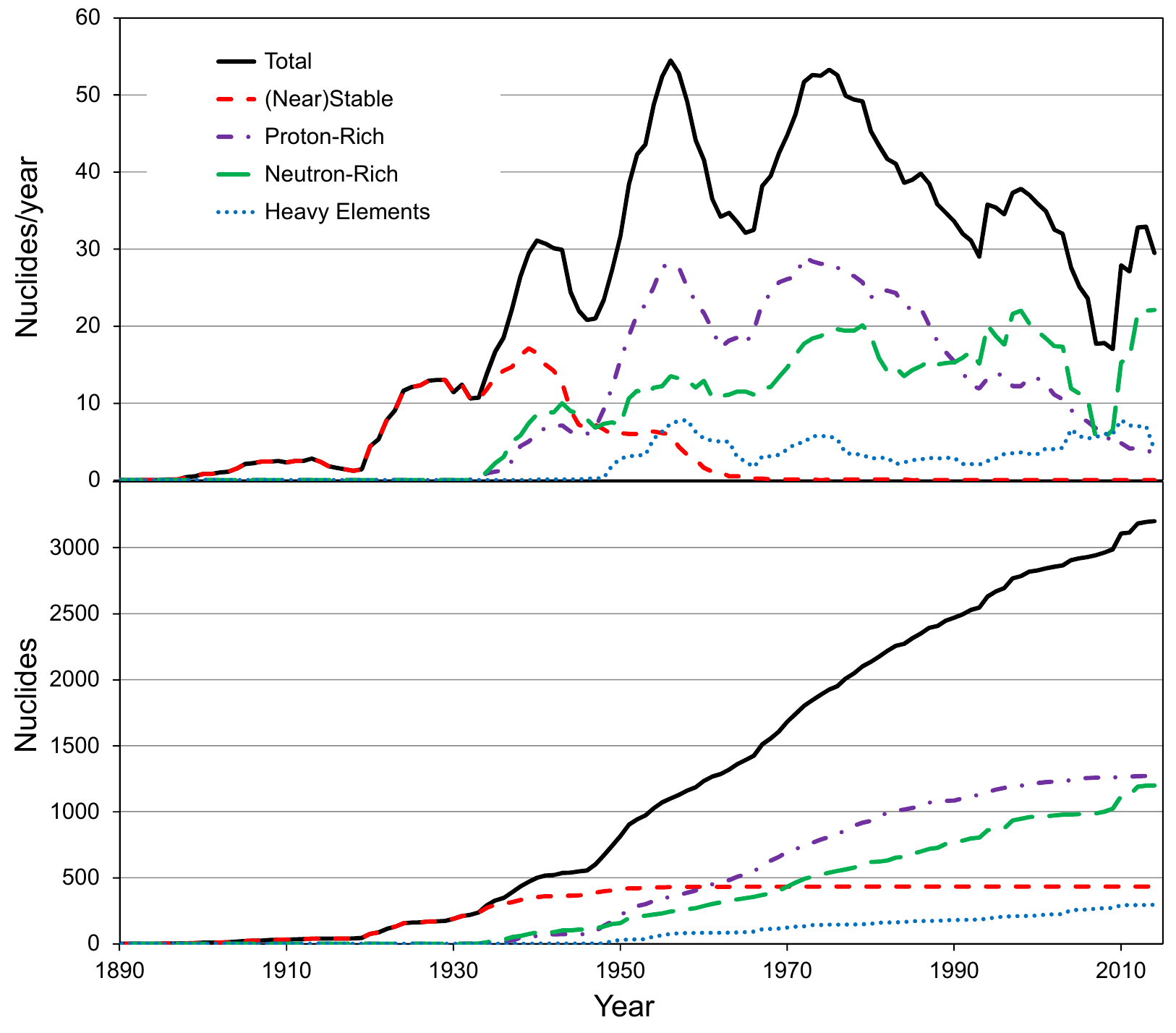,width=10.8cm}}
\caption{Discovery of nuclides as a function of year. The top figure shows the 10-year running average of the number of nuclides discovered per year while the bottom figure shows the cumulative number.  The total number of nuclides shown by the black, solid lines are plotted separately for near-stable (red, short-dashed lines), neutron-deficient (purple, dot-dashed lines), neutron-rich (green, long-dashed lines) and transuranium (blue, dotted lines) nuclides. This figure was adapted from Ref. 5 to include the data from 2014. \label{f:timeline} }
\end{figure}

With the new discoveries in 2014 and the reassignments described above the current status of the evolution of the nuclide discovery is shown in Figure \ref{f:timeline}. The figure was adapted from the previous review\cite{2014Tho01} and was extended to include 2014. The top part of the figure shows the ten-year average of the number of nuclides discovered per year while the bottom panel shows the integral number of nuclides discovered. It can be seen that the recent rate increase that started in 2010 did not continue and the rate dropped again below 30 to 29.5 nuclides/year.  

Even though last year only two new neutron-rich nuclides were discovered the ten-year average rate still increased slightly from 22.0 to 22.1 corresponding to the largest rate for the discovery of neutron-rich nuclides ever. Still the total number of known neutron-rich nuclides is still smaller (1199) than the total number of proton-rich nuclides (1274).

The rate of new proton-rich nuclides continues to decline reaching the smallest number since 1937 with 3.5 nuclides/year. Also the discovery rate of heavy-element nuclides dropped significantly to 3.9 corresponding to the lowest number in 15 years.

Overall 3191 different nuclides have been discovered so far (the number of 3195 quoted in the previous update \cite{2014Tho01} counted the 10 isotopes that have been discovered simultaneously by two different groups twice and should have been 3185). The discoveries were reported by 901 different first authors in 1527 papers and a total of 3546 different coauthors.

\begin{table}[pt] 
\tbl{Top ten countries where the most nuclides were discovered. The total number of nuclides are listed together with first and most recent year of a discovery. \label{countries}}
{\begin{tabular}{@{}rlrcc@{}} \toprule
Rank & Country & Number & First year & Recent year \\ \colrule
1	&	 USA 		&	1326	&	1907	&	2013	\\
2	&	 Germany 	&	557	&	1898	&	2014	\\
3	&	 UK 		&	300	&	1900	&	1994	\\
4	&	 Russia 	&	248	&	1957	&	2013	\\
5	&	 France 	&	217	&	1896	&	2005	\\
6	&	 Switzerland 	&	128	&	1934	&	2009	\\
	&	 Japan 	&	128	&	1938	&	2014	\\
8	&	 Canada 	&	62	&	1945	&	1993	\\
9	&	 Sweden 	&	60	&	1900	&	1998	\\
10	&	 Finland 	&	39	&	1961	&	2014	\\
\botrule
\end{tabular}}
\end{table}

Table \ref{countries} lists the 10 countries where the most nuclides were discovered. Overall nuclides were discovered in 25 different countries. There were essentially no changes over the last three years \cite{2012Tho03,2013Tho02,2014Tho01}. As before, the numbers for  Germany and Russia include nuclides discovered in West-Germany between 1949 and 1990 and the USSR from 1957 through 1991, respectively. Nuclides that were discovered simultaneously in two different countries are counted for each country so that the total sum is 3204 instead of 3191. Japan is now tied with Switzerland for sixth and the reassignments of the discovery of $^{80}$Zn from Sweden to the USA and $^{71}$Br from Switzerland to Canada dropped Sweden to ninth behind Canada at eighth. 

Over 120 different laboratories reported new nuclides. The top ten laboratories are listed in Table \ref{labs}. The only major difference from last year is the move of RIKEN from tenth to seventh which is mostly due to the reassignment of the discoveries in the 1930s and 1940s from Tokyo to RIKEN.

\begin{table}[pt] 
\tbl{Top ten laboratories where the most nuclides were discovered. The total number of nuclides are listed together with the country of the laboratory and the first and most recent year of a discovery. \label{labs}}
{\begin{tabular}{@{}rllrcc@{}} \toprule
Rank & Laboratory & Country & Number & First year & Recent year \\ \colrule
1	&	 Berkeley 	& USA		&	635	&	1928	&	2010	\\
2	&	 GSI	 	& Germany	&	437	&	1977	&	2014	\\
3	&	 Dubna 	& Russia	&	223	&	1957	&	2013	\\
4	&	 Cambridge 	& UK		&	222	&	1913	&	1940	\\
5	&	 CERN 	& Switzerland&	118	&	1965	&	2009	\\
6	&	 Argonne 	& USA		&	117	&	1947	&	2012	\\
7	&	 RIKEN 	& Japan	&	86	&	1972	&	2014	\\
8	&	 GANIL 	& France	&	85	&	1985	&	2005	\\
9	&	 Oak Ridge 	& USA		&	77	&	1946	&	2006	\\
10	&	 Orsay 	& France	&	73	&	1959	&	1989	\\
	&	 Michigan State 	& USA &	73	&	1967	&	2013	\\

\botrule
\end{tabular}}
\end{table}

The last update\cite{2014Tho01} included additional tables for the top first authors, coauthors, production methods, and journals where the discoveries were reported. These lists are not included in the present update because there were no major changes relative to last year. Overview tables of the various demographics of the nuclide discoveries from all previous years are available on the internet.\cite{2011Tho03}

\section{Discoveries not yet published in refereed journals}

\begin{table}[pt]
\tbl{Nuclides only reported in proceedings or internal reports until the end of 2014. The nuclide, first author, year, laboratory, conference or report and reference of the discovery are listed. \label{reports}}
{\begin{tabular}{@{}lllllr@{}} \toprule
\parbox[t]{1.5cm}{\raggedright Nuclide(s) \vspace*{0.2cm}}& Author & Year & Lab. & Conf./Report & Ref.\\ \colrule
\parbox[t]{1.5cm}{\raggedright $^{81,82}$Mo, $^{85,86}$Ru \vspace*{0.2cm}}&  H. Suzuki et al.  & 2013 &  RIKEN  &  \parbox[t]{2.8cm}{\raggedright EMIS 2012 \vspace*{0.2cm}}  &  \refcite{2013Suz01} \\$^{156}$Pr&  S. Czajkowski et al.  & 1996 &  GSI  &  \parbox[t]{2.8cm}{\raggedright ENAM'95  \vspace*{0.2cm}}  &  \refcite{1996Cza01} \\
\parbox[t]{1.5cm}{\raggedright $^{126}$Nd, $^{136}$Gd, $^{138}$Tb  \vspace*{0.2cm}}&  G. A. Souliotis  & 2000 &  MSU  &  \parbox[t]{2.8cm}{\raggedright Achiev. and Persp. in Nucl. Struct. 1999 \vspace*{0.2cm}}  &  \refcite{2000Sou01} \\
\parbox[t]{1.5cm}{\raggedright$^{143}$Ho \vspace*{0.2cm}}  &  G. A. Souliotis  & 2000 &  MSU  &  \parbox[t]{2.8cm}{\raggedright Achiev. and Persp. in Nucl. Struct. 1999 \vspace*{0.08cm}}  &  \refcite{2000Sou01} \\
  &  D. Seweryniak et al.  & 2002 &  LBL  &  \parbox[t]{2.8cm}{\raggedright Annual Report \vspace*{0.2cm}}  &  \refcite{2003Sew02} \\
\parbox[t]{1.5cm}{\raggedright$^{144}$Tm \vspace*{0.05cm}}  &  K. P. Rykaczewski et al.  & 2004 &  ORNL  &  \parbox[t]{3.4cm}{\raggedright Nuclei at the Limits 2004  }  &  \refcite{2005Ryk01} \\
 &  R. Grzywacz et al.  &   &   &  \parbox[t]{2.8cm}{\raggedright ENAM2004 \vspace*{0.05cm}}  &  \refcite{2005Grz01} \\
 &  C. R. Bingham et al.  &   &   &  \parbox[t]{2.8cm}{\raggedright CAARI2004 \vspace*{0.2cm}}  &  \refcite{2005Bin01} \\
\parbox[t]{1.5cm}{\raggedright $^{150}$Yb, $^{153}$Hf   \vspace*{0.2cm}} &  G. A. Souliotis  & 2000 &  MSU  &  \parbox[t]{2.8cm}{\raggedright Achiev. and Persp. in Nucl. Struct. 1999 \vspace*{0.2cm}}  &  \refcite{2000Sou01} \\
\parbox[t]{1.5cm}{\raggedright$^{234}$Cm \vspace*{0.08cm}}  &  P. Cardaja et al.  & 2002 &  GSI  &  \parbox[t]{2.8cm}{\raggedright Annual Report \vspace*{0.05cm}}  &  \refcite{2002Cag01} \\
 &  J. Khuyagbaatar et al.  & 2007 &  GSI  &  \parbox[t]{2.8cm}{\raggedright Annual Report \vspace*{0.05cm}}  &  \refcite{2007Khu01} \\
 &  D. Kaji et al.  & 2010 &  RIKEN  &  \parbox[t]{2.8cm}{\raggedright Annual Report \vspace*{0.2cm}}  &  \refcite{2010Kaj01} \\
\parbox[t]{1.5cm}{\raggedright$^{235}$Cm \vspace*{0.2cm}}  &  J. Khuyagbaatar et al.  & 2007 &  GSI  &  \parbox[t]{2.8cm}{\raggedright Annual Report \vspace*{0.2cm}}  &  \refcite{2007Khu01} \\
\parbox[t]{1.5cm}{\raggedright$^{234}$Bk \vspace*{0.2cm}}  &  K. Morita et al.  & 2002 &  RIKEN  &  \parbox[t]{2.8cm}{\raggedright Front. of Coll. Motion 2002 \vspace*{0.1cm}}  &  \refcite{2003Mor02} \\
 &  K. Morimoto et al.  &   &   &  \parbox[t]{2.8cm}{\raggedright Annual Report \vspace*{0.05cm}}  &  \refcite{2003Mor01} \\
 &  D. Kaji et al.  & 2010 &  RIKEN  &  \parbox[t]{2.8cm}{\raggedright Annual Report \vspace*{0.2cm}}  &  \refcite{2010Kaj01} \\
\parbox[t]{1.5cm}{\raggedright$^{252,253}$Bk \vspace*{0.2cm}}  &  S. A. Kreek et al.  & 1992 &  LBL  &  \parbox[t]{2.8cm}{\raggedright Annual Report \vspace*{0.2cm} }  &  \refcite{1992Kre01} \\
\parbox[t]{1.5cm}{\raggedright$^{262}$No \vspace*{0.08cm}}  &  R. W. Lougheed et al.  & 1988 &  LBL  &  \parbox[t]{2.8cm}{\raggedright Annual Report }  &  \refcite{1988Lou01} \\
 &   &   &   &  \parbox[t]{3.4cm}{\raggedright 50 Years Nucl.\ Fiss.\ 1989 \vspace*{0.08cm}}  &  \refcite{1989Lou01} \\
 &  E. K. Hulet  &   &   &  \parbox[t]{2.8cm}{\raggedright Internal Report \vspace*{0.2cm} }  &  \refcite{1989Hul01} \\
\parbox[t]{1.5cm}{\raggedright$^{261}$Lr  \vspace*{0.08cm}} &  R. W. Lougheed et al.  & 1987 &  LBL  &  \parbox[t]{2.8cm}{\raggedright Annual Report \vspace*{0.05cm}}  &  \refcite{1987Lou01} \\
 &  E. K. Hulet  &   &   &  \parbox[t]{2.8cm}{\raggedright Internal Report \vspace*{0.05cm}}  &  \refcite{1989Hul01} \\
 &  R. A. Henderson et al.  & 1991 &  LBL  &  \parbox[t]{2.8cm}{\raggedright Annual Report \vspace*{0.2cm}}  &  \refcite{1991Hen01} \\
\parbox[t]{1.5cm}{\raggedright$^{262}$Lr \vspace*{0.08cm}}  &  R. W. Lougheed et al.  & 1987 &  LBL  &  \parbox[t]{2.8cm}{\raggedright Annual Report \vspace*{0.05cm}}  &  \refcite{1987Lou01} \\
 &  E. K. Hulet  &   &   &  \parbox[t]{2.8cm}{\raggedright Internal Report \vspace*{0.05cm}}  &  \refcite{1989Hul01} \\
 &  R. A. Henderson et al.  & 1991 &  LBL  &  \parbox[t]{2.8cm}{\raggedright Annual Report \vspace*{0.2cm}}  &  \refcite{1991Hen01} \\
\parbox[t]{1.5cm}{\raggedright$^{255}$Db  \vspace*{0.08cm}} &  G. N. Flerov  & 1976 &  Dubna  &  \parbox[t]{2.8cm}{\raggedright  Nuclei Far from Stability 1976 \vspace*{0.08cm}}  &  \refcite{1976Fle01} \\ 
\botrule
\end{tabular}}
\end{table}

In last year's update\cite{2014Tho01} a table which listed all nuclides that had only been reported in conference proceedings was presented. Only one of the nuclides listed in the table ($^{164}$Ir) has been published in a refereed journal in 2014. No new discoveries were reported in conference proceedings during the last year. Thus, Table \ref{reports} is essentially identical to the table included in Ref. \refcite{2014Tho01}. It is reprinted here for completeness and as an encouragement to verify the existence of these nuclides by publishing the results in a refereed journal.

\section{Outlook for 2015}
The relatively small number of new nuclides discovered in 2014 had been predicted last year\cite{2014Tho01} because only RIBF\cite{2010Sak01} at RIKEN is in the position to produce a large number of new nuclides within a single experiment. The other two new next generation fragmentation facilities - FAIR\cite{2009FAI01} at GSI and FRIB\cite{2010Bol01,2012Wei01} at MSU - will not be ready until at least the end of the decade.

Only one nuclide mentioned in last year's outlook ($^{205}$Ac) \cite{2014Zha01} was discovered in 2014. The thirteen neutron-rich nuclides ($^{153}$Ba, $^{154,155}$La, $^{156,157}$Ce, $^{156-160}$Pr, $^{162}$Nd, $^{164}$Pm, and $^{166}$Sm) that  Kubo presented at the 2013 International Nuclear Physics Conference\cite{2013Kub01} have not been published yet. Also, $^{20}$B and $^{21}$C were only presented at the second Conference on Advances in Radioactive Isotope Science (ARIS).\cite{2014Mar01,2014Kon01} 
Other nuclides presented at ARIS were $^{215,216}$U, \cite{2014Wak01} $^{280}$Ds, \cite{2014Mor01} and $^{284}$Fl.\cite{2014Ryk01}

In addition, a large number of new nuclides were shown for the first time at two other conferences. Kameda presented the discovery of sixteen new nuclei ($^{156}$La, $^{158}$Ce, $^{161}$Pr, $^{163}$Nd, $^{165}$Pm, $^{167}$Sm, $^{169}$Eu, $^{171}$Gd, $^{173,174}$Tb, $^{175,176}$Dy, $^{177,178}$Ho, and $^{179,180}$Er) at the 16$^{th}$ ASRC International Workshop on Nuclear Fission and Structure of Exotic Nuclei in Tokai\cite{2014Kam01} and Shimizu presented another twenty-five nuclei ($^{116}$Nb, $^{118}$Mo, $^{121,122}$Tc, $^{125}$Ru, $^{127}$Rh, $^{129,130,131}$Pd, $^{132}$Ag, $^{134}$Cd, $^{136,137,138}$In, $^{139,140}$Sn, $^{141,142,143}$Sb, $^{144,145}$Te $^{146,147}$I, $^{149}$Xe, and $^{154}$Ba) at the 4$^{th}$ Joint Meeting of the APS Division of Nuclear Physics and the Physical Society of Japan.\cite{2014Shi01}

Based on these presentations, it is not unreasonable to expect that at least thirty new nuclides should be published during 2015.

\section{Summary}
The documentation of the history of the nuclide discoveries remains a work in progress as evidenced by the continued revisions. The original project extended over a period of four years between 2007 and 2011. During this time the criteria evolved and thus they were not equally applied to all nuclides. Currently, the discoveries of all nuclides is being reviewed in order to apply established criteria equally and consistently. Thus, further revisions are to be expected. Input from other researchers in this process is encouraged and readers should contact the author with corrections and suggestions. 

\section*{Acknowledgements}

I would like to thank Professor Brett Thornton for pointing out the work by Hulubei and Cauchois on the discovery of astatine and Professor Hide Sakai for discussions of the history of nuclide discoveries at RIKEN. I also would like to thank Ute Thoennessen for carefully proofreading the manuscript. Support of the National Science Foundation under grant No. PHY11-02511 is gratefully acknowledged.

\vfill
\newpage

\bibliographystyle{ws-ijmpe}
\bibliography{../springer-book/tex-file/isotope-references}

\end{document}